\documentclass[fleqn,10pt]{wlscirep}
\usepackage{textcomp}
\title{Experimental demonstration of negative refraction with 3D locally resonant acoustic metafluids}

\author[1]{Benoit Tallon}
\author[2]{Artem Kovalenko}
\author[1]{Olivier Poncelet}
\author[1]{Christophe Arist\'egui}
\author[2]{Olivier Mondain-Monval}
\author[1,*]{Thomas Brunet}
\affil[1]{Univ. Bordeaux, CNRS, Bordeaux INP, ENSAM, I2M, UMR 5295, F-33405, Talence, France}
\affil[2]{Univ. Bordeaux, CNRS, CRPP, F-33600, Pessac, France}

\affil[*]{thomas.brunet@u-bordeaux.fr}

\begin{abstract}
Negative refraction of acoustic waves is demonstrated through underwater experiments conducted at ultrasonic frequencies on a 3D locally resonant acoustic metafluid made of soft porous silicone-rubber micro-beads suspended in a yield-stress fluid. By measuring the refracted angle of the acoustic beam transmitted through this metafluid shaped as a prism, we determine the acoustic index to water according to Snell's law. These experimental data are then compared with an excellent agreement to calculations performed in the framework of Multiple Scattering Theory showing that the emergence of negative refraction depends on the volume fraction $\Phi$ of the resonant micro-beads. For diluted metafluid ($\Phi=3\%$), only positive refraction occurs whereas negative refraction is demonstrated over a broad frequency band with concentrated metafluid ($\Phi=17\%$).
\end{abstract}
\begin{document}

\flushbottom
\maketitle

\thispagestyle{empty}

\section*{Introduction}

Since the pioneering works reported by Liu \textit{et al.}\cite{Liu2000_289_1734}, locally resonant acoustic metamaterials have been attracting great attention \cite{Ma2016_2_e1501595}. One of the challenging issues has been the achievement of acoustic metamaterials with a negative refractive index that offer new possibilities for acoustic imaging materials and for the control of sound at sub-wavelength scales \cite{Cummer2016_1_16001}. In acoustics, the refractive index $n$ is proportional to $\sqrt{\rho/K}$, where $K$ and $\rho$ are the bulk modulus and the mass density of the material. Many works have been devoted to the study of double-negative metamaterials \cite{Li2004_70_055602,Lee2010_104_054301,Liang2012_2_859,Yang2013_110_134301,Maurya2016_6_33683,Lanoy2017_22_220201,Zhou2018_10_044006,Wang2018_10_064011,Dong2020_137_103889} for which the two constitutive parameters $K$ and $\rho$ are simultaneously negative, leading thus to a negative index\cite{Li2007}. It is worth noting that such a double-negativity condition is not required for (real) dissipative metamaterials to get a negative index\cite{Brunet2015_2_3}. When they are non-negligible, losses may play an important role in the effective acoustic properties of the metamaterials in such a way that the real part of the acoustic index can be negative for single-negative metamaterial as demonstrated in underwater ultrasonic experiments\cite{Brunet2015_14_384}, and in air at audible frequencies\cite{Kaina2015_525_77}. Although the latter work reported on the experimental observation of negative refraction effects within a 2D acoustic superlens, negative refraction has never been observed with a 3D acoustic metamaterial up to now.

In that context, we proposed to use a "soft" approach, combining various soft-matter techniques, to achieve soft 3D acoustic metamaterials with negative index composed of resonant porous micro-beads randomly-dispersed in a yield-stress fluid\cite{Brunet2013_342_323}. By taking benefit from the strong low-frequency Mie-type (monopolar and dipolar) resonances of these 'ultra-slow' particles, single-band\cite{Brunet2015_14_384} and dual-band\cite{Raffy2016_28_1760} negative refractive indices were experimentally demonstrated. The issue of the experimental observation of negative refraction was then raised for these water-based metamaterials\cite{Popa2015_14_363}, since the energy attenuation might be significant in these metafluids due to the intrinsic absorption in the porous micro-beads, and to the strong resonant scattering by the particles.

\begin{figure}[htbp]
\begin{center}
	\includegraphics[width=11cm]{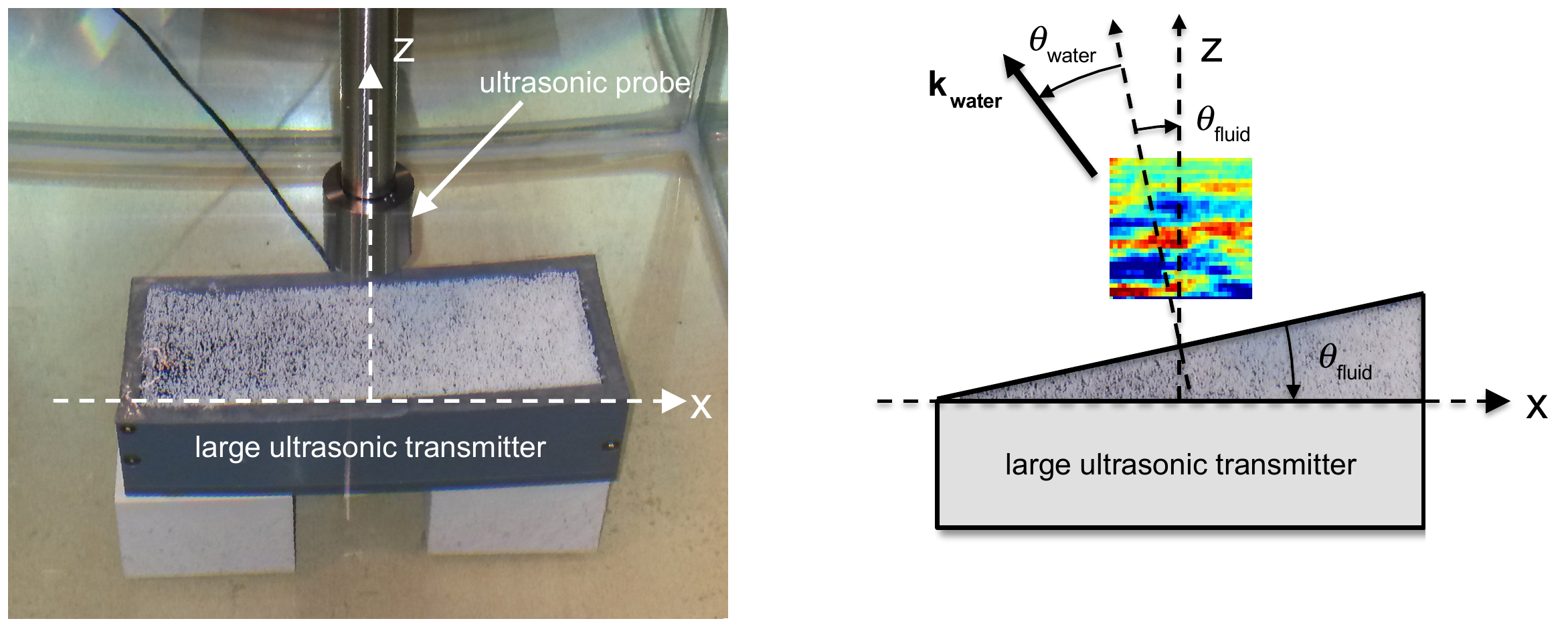}
	\caption{(left) Top view and (right) side view of the experimental setup. The metafluid is confined in a prism-shaped box, with an angle $\theta_{\rm{fluid}}=+2$\textdegree, that is deposited directly on a large broad-band ultrasonic immersion transducer (150 mm x 40 mm). The acoustic waves propagate from the bottom to the top of the water tank, along the vertical $z$-axis. The acoustic field refracted from the prism is scanned in the $x$-$z$ plane (33 mm x 33 mm) with a small ultrasonic probe.}
	\label{figure1}
\end{center}
\end{figure}

In this paper, we report on negative refraction experiments with metafluids, composed of soft porous silicone-rubber micro-beads, exhibiting a negative acoustic index at ultrasonic frequencies\cite{Brunet2015_14_384}. In these experiments, the metafluid is confined in a prism-shaped box with a small angle ($\theta_{\rm{fluid}}=+2$\textdegree) and with a very thin plastic surface in order to be acoustically transparent. As shown in Fig. \ref{figure1}, this metafluid is directly deposited on a large ultrasonic immersion transducer operating in water over a broad ultrasonic frequency range (from 15 kHz to 600 kHz). The large dimensions of this transmitter ensures the generation of quasi-pure plane waves propagating in the metafluid along the vertical $z$-axis, from the bottom to the top. Then, the transmitted acoustic beam refracted by the interface metafluid/water is scanned in the water tank by using a small ultrasonic probe. Two samples, composed of micro-beads with similar mean diameters $d$ of about 750 $\mu$m, are considered in this study with two different volume fractions $\Phi$, referred to as the diluted metafluid ($\Phi=3\%$) and concentrated metafluid ($\Phi=17\%$). The acoustic properties of the micro-beads are given in a previous work\cite{Kovalenko2016_12_5154}. One of the particular features of these porous particles is their very low longitudinal phase velocity $c_L$ (= 120 m.s$^{-1}$) that is due to the softness of the silicone-rubber material\cite{Ba2017_7_40106}.

\section*{Results}

\begin{figure}[htbp]
\begin{center}
	\includegraphics[width=13.5cm]{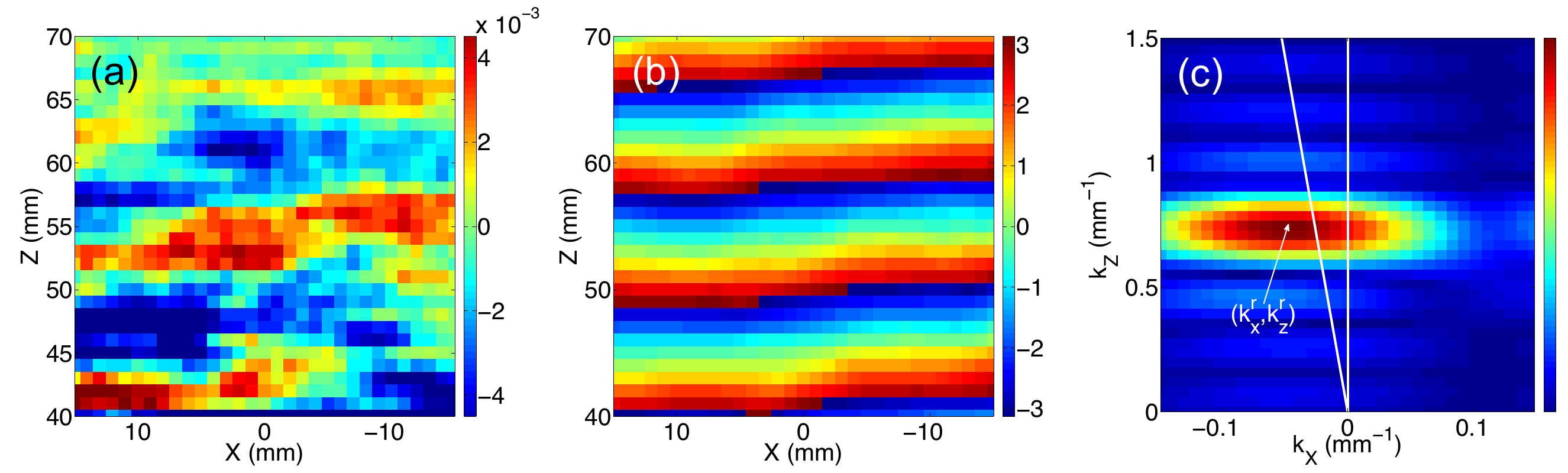}
	\caption{(a) Snapshot of the measured pressure field refracted from the prism filled with concentrated metafluid ($\Phi=17\%$). (b) Corresponding angular phase, shown here at 175 kHz, obtained from the time-domain Fourier transforms performed for each position over the square grid shown in (a). (c) 2D spatial Fourier transform of the scanned field pattern at 175 kHz for the measurement of the (real-valued) spatial Fourier components $k^{\textrm{r}}_{\textrm{x}}$ and $k^{\textrm{r}}_{\textrm{z}}$ of the refracted beam in water.}
	\label{figure2}
\end{center}
\end{figure}

When excited with a short electrical pulse, the large ultrasonic transducer generates a short broad-band acoustic pulse propagating in the confined fluid along the $z$-axis only. At the interface between the confined fluid and water, the acoustic pulse is then refracted in the surrounding water with an angle of refraction $\theta_{\rm{water}}$ shown in Fig. \ref{figure1}. By measuring $\theta_{\rm{water}}$, the acoustic index $n_{\rm{fluid}}$ of the fluid confined in the prism can be easily deduced from the Snell's law as following:

\begin{equation}
n_{\rm{fluid}} = n_{\rm{water}} \frac{\rm{sin}(\theta_{\rm{water}})}{\rm{sin}(\theta_{\rm{fluid}})}
\label{equation1}
\end{equation}

with $n_{\rm{water}}=1$ since the acoustic index $n$ ($=c_0/c$) of a material with the phase velocity $c$ is usually defined relatively to water ($c_0=c_{\textrm{water}}$) for underwater acoustics. In these experiments, the acoustic index $n_{\rm{fluid}}$ cannot be directly retrieved from the refracted transmitted temporal signals measured in the $x$-$z$ plane as depicted in Fig. \ref{figure1}, since a short broad-band acoustic pulse has been used in these pulsed ultrasonic experiments. In a such broad frequency range (from 15 kHz to 600 kHz), the acoustic index of a concentrated metafluid is expected to exhibit strong variations ranging from high positive values to negative ones\cite{Brunet2015_14_384}. Due to potential strong dispersion effects, we focus here on a harmonic spectral analysis of the refracted beams by performing Fourier transforms over time and space. First, we performed time-domain Fourier transforms of all the signals acquired at each position over a 2D spatial grid (Fig. \ref{figure2}.a.) in order to obtain the spatial-field at each frequency component (Fig. \ref{figure2}.b.). Then a 2D spatial Fourier transform is applied at each frequency for getting the wavenumber spectrum of the harmonic beam (Fig. \ref{figure2}.c.). Since the beams refracted in water are quasi-plane waves, it is straightforward to extract accurately their direction of propagation from the (real-valued) spatial Fourier components $k^{\textrm{r}}_{\textrm{x}}$ and $k^{\textrm{r}}_{\textrm{z}}$ at the peak amplitude of the wavenumber spectrum.

\begin{figure}[htbp]
 \begin{center}
 	\includegraphics[width=16cm]{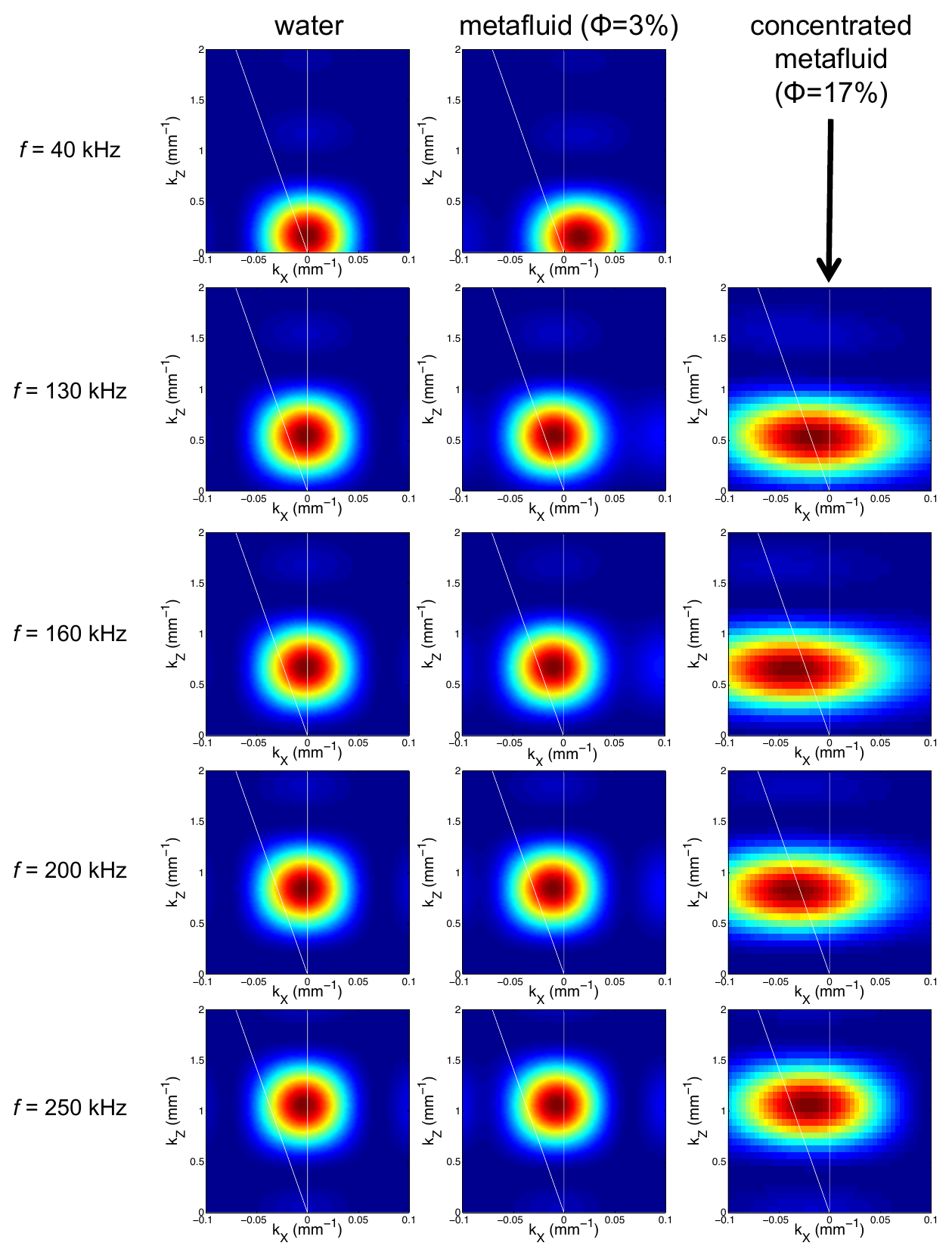}
	\caption{2D Fourier transforms of the scanned field pattern performed at different frequencies: 40, 130, 160, 200, 250 kHz (from the top to bottom). The prism is filled with either water (left), diluted metafluid (center) or concentrated metafluid (right). On all maps, the white vertical axis ($k_{\textrm{x}}=0$) corresponds to the case in which $\theta_{\textrm{water}}=\theta_{\textrm{fluid}}$ meaning that $n_{\textrm{fluid}}=+1$, whereas the tilted white axis ($\arctan(k_{\textrm{x}}/k_{\textrm{z}})=-\theta_{\rm{fluid}}$) corresponds to the case in which $\theta_{\textrm{water}}=0$ meaning that $n_{\textrm{fluid}}=0$. The Fourier transforms imply a plane wave in the positive $z$ direction.}
	\label{figure3}
\end{center}
\end{figure}

From the acoustic field map shown at a given time in Fig. \ref{figure2}.a, we can get the corresponding map of the angular phase for each frequency component of the refracted beam. As an example, Fig. \ref{figure2}.b shows the angular phase of the refracted beam at $f=175$ kHz revealing an angle of refraction of a few degrees from the initially vertical propagation direction. Such a deviation can be estimated by measuring the tilted angle of these refracted wavefronts in the $x$-$z$ plane but this direct spatial measurement may suffer from fluctuations observed on the wavefronts. As an alternative, we performed 2D Fourier transforms of the scanned field patterns at different frequencies to infer the $k^{\textrm{r}}_{\textrm{x}}$ and $k^{\textrm{r}}_{\textrm{z}}$ coordinates of the wave vector $\textbf{k}_{\textrm{water}}$ in the $k_\textrm{x}$-$k_\textrm{z}$ plane. The values of these two components are given by the location of the maximum spot intensity shown in Fig. \ref{figure2}.c. Then, the angle of refraction $\theta_{\rm{water}}$ can be easily deduced for each frequency as following:

\begin{equation}
\theta_{\rm{water}}=\arctan(\frac{k^{\textrm{r}}_{\textrm{x}}}{k^{\textrm{r}}_{\textrm{z}}})+\theta_{\rm{fluid}}
\label{equation2}
\end{equation}

In these experiments, the coordinate $k^{\textrm{r}}_{\textrm{z}}$ is necessarily positive since the acoustic waves propagate from the bottom to the top of water tank along the $z$-axis in any case. However, the coordinate $k^{\textrm{r}}_{\textrm{x}}$ may be either positive or negative depending on the direction of the refracted beam. The acoustic index $n_{\rm{fluid}}$ of the fluid confined in the prism can be deduced from the values of the angle $\theta_{\rm{water}}$ by using Eq. \ref{equation1}. When the prism is filled with water (Fig. \ref{figure3}, left), the spot goes along the vertical $k_{z}$-axis for which $k^{\textrm{r}}_{\textrm{x}}=0$ (white vertical line) as the frequency increases leading to $\theta_{\rm{water}}=\theta_{\rm{fluid}}$ according to Eq. \ref{equation2}. Therefore, no refraction occurs in that case which may be expected since the material is the same (water) on both sides of the interface. Note that if the spot had gone along the white titled line shown in Figs. \ref{figure3}, for which $\arctan(k^{\textrm{r}}_{\textrm{x}}/k^{\textrm{r}}_{\textrm{z}})=-\theta_{\textrm{fluid}}$, this would lead to $\theta_{\textrm{water}}=0$, corresponding thus to a zero-index fluid confined in the prism.

\begin{figure*}[htbp]
 \begin{center}
	\includegraphics[width=10cm]{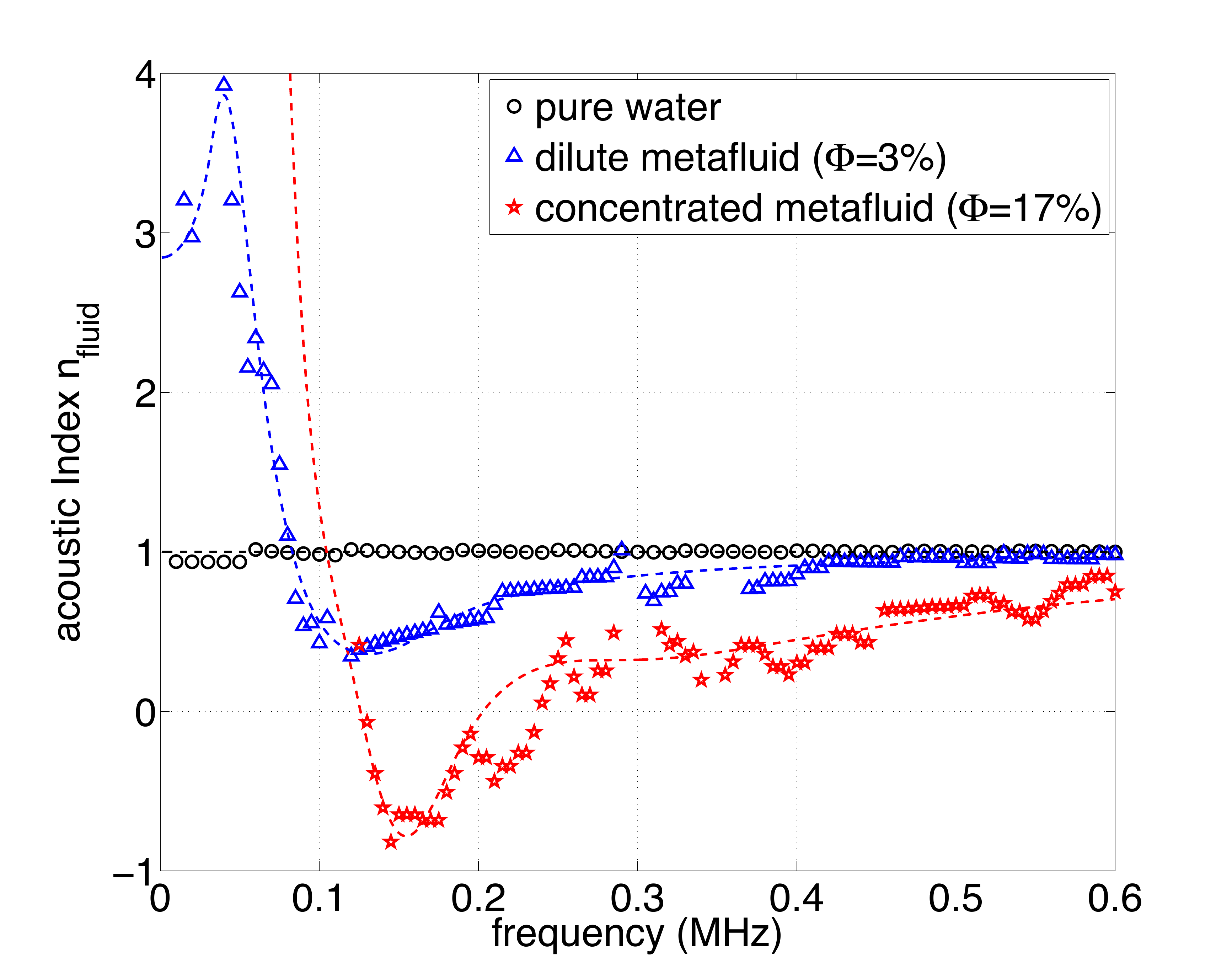}
	\caption{Acoustic index $n_{\textrm{fluid}}$ for different fluids confined in the prism-shaped box and extracted as a function of frequency from the $k$-maps. Dashed lines refer to calculations performed in the framework of multiple scattering theory (see the text).}
	\label{figure4}
	\end{center}
\end{figure*}

When the prism is filled with the diluted metafluid ($\Phi=3\%$, $d=750$ $\mu$m with a size dispersion of 30\%), the spot oscillates around the white vertical axis as the frequency increases (Fig. \ref{figure3}, center). The values of $n_{\textrm{fluid}}$ extracted from Eqs.\ref{equation1} and \ref{equation2} show that the acoustic index of this diluted metafluid varies from high values at low frequencies ($n_{\textrm{fluid}}=+4$ at 40 kHz) to low values at intermediate frequencies ($n_{\textrm{fluid}}=+0.5$ around 130 kHz) before getting closer to +1 at high frequencies as shown in Fig. \ref{figure4}. This strong dispersion is due to the micro-bead resonances that occur around 150 kHz for that size of particles. Far away from these low-frequency acoustic resonances, the acoustic index of the metafluid is similar to that of the aqueous matrix\cite{Brunet2015_14_384}.

We also produced a concentrated metafluid ($\Phi=17\%$, $d=700$ $\mu$m with a size dispersion of 10\%) in which negative refraction occurs for a certain range of frequencies. Actually, Fig. \ref{figure3} (right) shows that the spot can go beyond the white tilted axis (corresponding to $n_{\textrm{fluid}}=0$) as observed at 160 kHz. The experimental results are not shown at 40 kHz because of the very high attenuation that is due the strong (monopolar) low-frequency resonances of the micro-beads \cite{Brunet2015_14_384}. But the acoustic index is shown to be negative over a broad frequency range as shown in Fig. \ref{figure4}. Note that this 'negative band' is slightly shifted to higher frequencies compared to the diluted metafluid because of the smaller size of the micro-beads of this concentrated metafluid \cite{Brunet2012_101_011913}.

Finally, we compared our acoustical measurements to theoretical predictions produced through multiple-scattering modeling, revealing good qualitative agreement as shown in Fig. \ref{figure4}. The theoretical acoustic index of the metafluids  ($\Phi=3\%$ and $17\%$) are obtained from their effective wavenumbers calculated using the Waterman-Truell formula\cite{Waterman1961_2_512}. The values of the material parameters for the soft porous silicone rubber that we used for these calculations, were $c_L$ = 120 m.s$^{-1}$, $\alpha_L$= 20 Np.m$^{-1}$.MHz$^{-2}$ (the phase velocity and attenuation coefficient for longitudinal waves), $c_T$ = 40 m.s$^{-1}$, $\alpha_T$= 200 Np.m$^{-1}$.MHz$^{-2}$ (the phase velocity and attenuation coefficient for shear waves) and $\rho_{1}$ = 760 kg.m$^{-3}$. The water-based gel matrix has the same properties as water ($\rho_0$ = 1000 kg.m$^{-3}$, $c_0$ = 1490 m.s$^{-1}$) since this host Bingham fluid is essentially made of water and very small amounts of polymer (Carbopol) to prevent the creaming of the porous micro-beads.

\section*{Conclusion}

In summary, we have reported an experimental demonstration of negative refraction in a 3D acoustic metamaterial. The experiments have been conducted at ultrasonic frequencies with a locally resonant metafluid composed of soft porous silicone-rubber micro-beads whose concentration must be high enough so that refraction be negative. However, the achievement of acoustic devices based on negative refraction such as perfect lenses envisioned by Pendry\cite{Pendry2000_85_3966} seems unattainable with our metafluids because of their large attenuation that is mainly due to the strong resonant scattering\cite{Brunet2015_14_384}. Alternatively to bulky and lossy 3D metamaterials, acoustic metasurfaces\cite{Assouar2018_3_460} might be much more appropriate to manipulate acoustic wavefronts as recently demonstrated with soft gradient-index metasurfaces\cite{Jin2019_10_143}.

\section*{Methods}

In this study, we used a large ultrasonic transducer (150 mm x 40 mm), with a central frequency of 150 kHz, to ensure the generation of quasi-pure plane waves in the water tank (propagating along the vertical axis from the bottom to the top of the water tank). This broad-band transducer was excited with a short electrical pulse generated by a pulser/receiver (Olympus, 5077PR) that was also used to amplify the electric signal recorded by the 1-inch-diameter receiving transducer (Olympus V301) before its acquisition on a computer via a waveform digitizer card (AlazarTech, ATS460). The angle of the prism has been chosen here as low as possible ($\theta_{\rm{fluid}}=+2$\textdegree) in order to guarantee that the refracted-beam amplitude does not vary too much along the x-axis in spite of the slightly increasing depth of the prism. For the experiments conducted with the prism filled with water (Fig. \ref{figure3}, left) or with the diluted metafluid (Fig. \ref{figure3}, centre), the width of the scan area along the x-axis was 60 mm. For the concentrated metafluid (Fig. \ref{figure3}, right), this width was reduced to 30 mm inducing a slight spread of the Fourier Transform distribution along the $k_x$-axis. The acoustical fields refracted in water were scanned on grids with a step of 1 mm that is 3 times smaller than the acoustic wavelength in water at 500 kHz (= 3 mm). Note that we also used zero-padding techniques before doing Fast Fourier Transforms in space to improve the resolution in the $k$-maps shown in Fig. \ref{figure3}.

\bibliography{biblio}

\section*{Acknowledgements}

This work was partially funded and performed within the framework of the Labex AMADEUS ANR-10-LABEX-0042-AMADEUS with the help of the French state Initiative d'Excellence IdEx ANR-10-IDEX-003-02 and project BRENNUS ANR-15-CE08-0024 (ANR and FRAE funds).

\section*{Author contributions statement}

T.B. supervised the project,  A.K. produced the soft porous silicone-rubber micro-beads to achieve the metafluids under the guidance of O.M.-M., B.T. and T.B. conducted the underwater experiments, B.T., T.B., C.A., and O.P analyzed the results. All authors reviewed the manuscript. 

\section*{Additional information}

\textbf{Competing financial interests} The authors declare no competing financial interests.

\end{document}